# An Approach to Log Management: Prototyping a Design of Agent for Log Harvesting.

**Reinaldo N. Mayol Arnao**

**<reinaldo.mayol@upb.edu.co>**

Universidad Pontificia Bolivariana y Universidad Industrial de Santander, Colombia

**Luis A. Núñez**

**<lnunez@uis.edu.co>**

Universidad Industrial de Santander, Colombia

**Antonio Lobo**

**<alobo@uis.edu.co>**

Universidad Industrial de Santander, Colombia

## Acknowledgments

The authors gratefully acknowledge the financial support of the Vicerrectoría de Investigación y Extensión de la Universidad Industrial de Santander under project GridUIS2 5541

## 1. Abstract

This paper describes a work in progress implementing a solution for harvesting and transporting information logs from network devices in a e-science environment. The system is composed for servers, agents, active devices and a transporting protocol.

This document describes the state of development of agents. Agents capture logs from devices, normalize, reduce and cataloged them by using metadata. Once all these processes are done, they transmit the cataloged data by using Transportation Protocol to a warehouse server. Also an agent use orchestration parameters to transmit modified logs to a data warehouse server. These parameters can be received from orchestration applications such as *Taverna*.

The operation of the agents and the communication protocol solve some of the deficiencies of traditional logs management protocols. Finally, we show some test realized over the new prototype.

## 2. Introduction

The traditional log management process includes several stages. The process begins when an application (for example a firewall daemon) generates an event (for example a IP packet that has been rejected) and writes its description, using an application[1] log format. In the second phase, if a log harvest system has been configured, the system log generates a network message for each event described in logs. This message is sent to storage a warehouse server. If the log harvest system is not set, logs are maintained local and it cannot be used for any type of process out of its environment. Usually in a production device, there are not possibilities for analysis, so in the best case, simple procedures of organizing information are made.

Important information concerning of networking management strategies is recorded in logs of most of the network devices. Today it is

---

[1] For simplicity the term applications is used but could refer to processes that do not correspond (as in the example) to the application layer of TCP / IP model (or OSI).

indispensable to obtain and process this information using sophisticated data mining techniques looking for patterns, trends or even, to detect specific acts that, in a certain context, can be considered as evidences for insecurity or malfunction. To accomplish it, with current conventional protocols[2] and working environments, have serious difficulties, among which is possible to mention:[1][2]

- The great diversity of security and telecommunication devices i.e., routers, firewalls, intruder detectors, antivirus systems, applications servers, among others.

- There is not one an accepted standard for the generation of logs. Each manufacturer (or even each model) generates information logs in different formats. For example, the IBM Autonomic Toolkit has defined 280 logs different formats. [3]

- Volumes of security and network operation information are very large, for example tracking activity of a local network can generate several gigabits of information per day.[4] [5]

- Volumes of scientific and commercial data have exploited. Today the capacities to mine data are lower than the posibility to generate them, and security incidents related have also increased exponentially.[6] [7]

- The computational power for management and analysis of logs (especially for correlational analysis) is very demanding. This makes it very difficult to obtain various "points of view" of information coming from logs, which is necessary for its correct analysis and interpretation of the information.

- The log information is perishable and volatile. Its preservation can be affected by its lifecycle and the storage and computing capacities of the corresponding devices that generate log.[8]

- Finally, the general accepted log transport mechanism (using Syslog daemon over UCP or TCP connections) has serious limits from performance, security, orchestration and mobile device identification.

Current most popular protocol used for harvest-transportation is Syslog[13]. In [15][16][5] are analyzed some limitations of this protocol. The most important limitations are:

1. It does not mechanisms to ensure:

    a. Data integrity.[3]
    b. Confidentiality of information.
    c. Authenticity at the ends of the connection.
    d. Clock synchronism among information sources.
    e. Time stamping of logs.
    f. Fault tolerance

2. The system is vulnerable to denial of service attacks, because one message is transmitted by each event (and no matter if this is repeated).

3. The system does not use a data compression mechanism.

4. The system cannot be orchestrated[4].

5. A single standard log format is not supported. This limits the real possibilities of data analysis.

6. The device identification is made by precarious mechanisms such as IP addresses or fully qualified domain names. This mechanism is not acceptable for mobile devices that can change its IP and generally do not have fully qualified domain names.

---

[2] Syslog, UDP and TCP.

[3] Limitations 1, 2 and 3 refer to the RFC described syslog protocol. There are some applications (for example rsyslog[66]) that have resolved this situation by using TLS.

[4] An orchestration model provides an environment specifically focused on the needs of a participant. These needs may change over time.

In [17](CLCAR 2010) we analyze some other limitations that appear when we tested Orchestration Applications to log management.

## 3. The proposed system description

In this work we present a system composed by Servers, Agents and Transportation Protocol. **Servers** storage information logs as syslog-like warehouse. **Agents** capture logs from devices, normalize, reduce and cataloged them by using metadata. Once all these processes are done, they transmit the cataloged data by using **Transportation Protocol**. The Figure 1 shows this process.

Along this paper, we shall describe the agents that :

- Harvest logs from network devices.
- Compress the information using frequency method.
- Send logs to a data warehouse through a new transportation protocol.
- Label the record with the appropriated metadata.
- Unifies logs format using Common Base Events (CBE) format.

### 3.1 Agents

An Agent is a software element that receives data logs from devices (local or remote) by using a TCP port, modify and transmit them to a data warehouse server. Agents use the difference between occurrence frequency expressions to reduce (compress) logs and normalize them to CBE format.

Agents are responsible of process for log from network devices and communication with servers, using transport protocol. The communication between the device and the agents occurs by a standard TCP port.

**Figure 1 The system proposed**

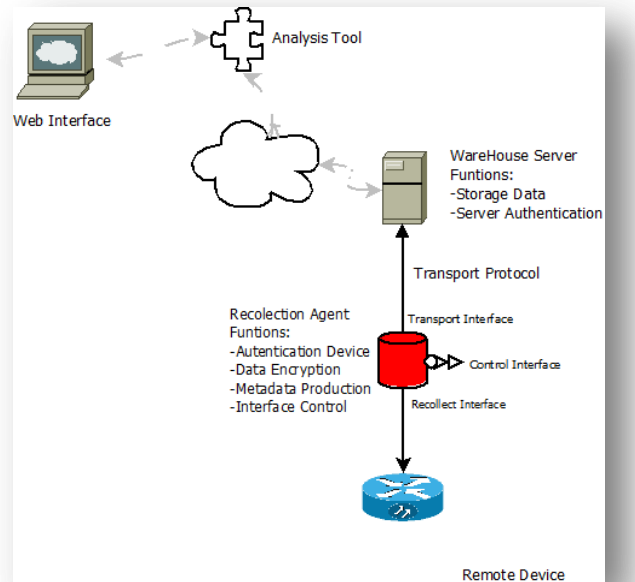

Each agent has three data interfaces:

- **Capturing Interface**: This interface interacts with the Device Log Systems, reads log lines and send them to a data warehouse.
- **Transporting Interface:** This interface is used for communication with the data warehouse server by using a transport protocol.
- **Controlling Interface:** This interface is used for agent operation and orchestration activity. It uses any TCP port according to the orchestration application settings.

Agents are responsible of important functions such as:

- **Data Compression:** The data compression will be performed by using a scanning frequency mechanism for regular expressions within the logs.[18] In a significant amount of logs, fixed expressions has high frequency of occurrence, for example:

    - Server ::: is down
    - User::: is connected in terminal :::
    - ::: interface is link down

    These are the *fixed expressions*.

Other expressions with a low occurrence frequencies correspond to particular names (user, daemons, interfaces) and are considered *variable expressions*. We use the occurrence frequency differences to establish a compression mechanism which records the line number and position (into log file) of each regular or variable expression. A similar mechanism (but only for knowing extract) is described in [18].

- **Data Normalization:** As it has been mentioned above that many log formats exist. Log format will be transformed CBE format. CBE is a common XML format that can be used for the representations of system events.[5]

- **Data Security:** Logs must be digitally signed before being transmitted. Additionally, the communication between Agents and Orchestration Application must be authenticated using Digital Certificates.

- **Digital Certificate Management:** Digital signatures comply X.509 v3 Digital Certificates. Agents store its Digital Certificates in a protected area. Protected Area is like a cryptographic key ring.

- **Control Interface:** The control interface is used to change some parameters for orchestration process, such as: server warehouse, kind of logs that must be recollected, cryptography control interface and others.

- **First Level of Metadata Characterization**: Two levels of metadata are established. The first level is used to describe time type of device and geographic location, data quality, harvest conditions and others parameters. The second level is used to replace raw data by identifiers constructed from frequency analysis.[13]

An agent use orchestration parameters to transmit modified logs to a data warehouse server. This transmission is made with a new Transportation Protocol which ensures that the information transmitted is cryptographically protected using Digital Certificates and both communication ends are authenticated.

The detailed explanation Transportation Protocol proposed is outside the scope of this paper. For this reason we show here only some important details. The transportation protocol is responsible for defining the communication among agents and warehouse servers and devices. This protocol must:

a) Use cryptographic authentication methods[5], instead of actual used IP address[6], as end identification system.

b) Support cryptographic methods to encrypt data logs transfer.

c) Offer clock synchronisms between agents, devices and servers.

d) Use a fixed header. Any other necessary information must be transmitted in an extension header.

The figure 2 shows the process for data (logs) interchange between an agent and a server using the transportation Protocol. As mentioned above the transportation protocol is the same between agents and servers or between agents and devices.

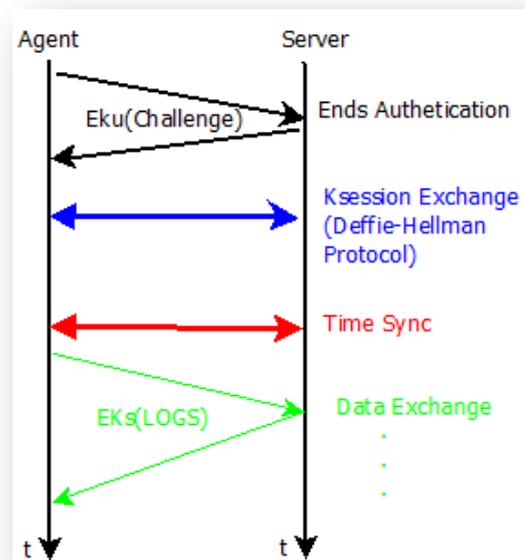

**Figure 2 Data Interchange between Agents and Servers. Ku represents the Public Key of Devices and Agents. Ks is the Session Key interchanged using Deffie-Hellman Protocol.**

---

[5] Digital Certificates

[6] Or Fully Qualified Domain Names

The figure 3 shows the structure of a message between agents and devices (or Servers in future). Agent ID is an identifier implemented by the agent digital certificate. Similarly, the Server ID is an identifier for this (and for other) type of device. The validation parameter defines the time until the message should be accepted by the agent. Message ID is a serial number used as identification in messages. Signature Protocol identifies the protocol used as hash function in Digital Signature, by default SHA-2. Encryption protocol defines the protocol used for data symmetric encryption; by default we have implemented AES. Message type specifies 128 types of message[7]. Finally, Message Signature is its digital signature. Message type must include the message priority like the syslog model, but using other message classification. [13]

**Figure 3 Message Protocol**

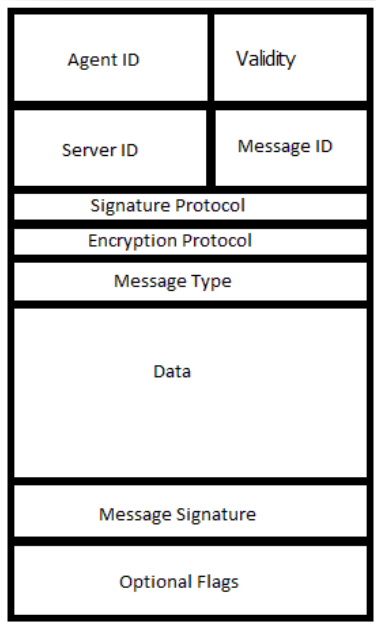

## 4. Current Status (Agents)

This section describes the current status of system developed. Presently agents are operational and the communication protocol (between devices and agents) has been partially accomplished. The communications between agents and devices has been successfully tested.

Agents are implemented by using Python script language and one exclusive threat is established for each device connected. Currently, devices begin the communication process, but in the near future agents will instruct the device to send specific information to the selected agent. Figure 3 shows an agent waiting for connections from devices. Figure 4 show 2 devices (in this case Linux machines) sending information to Agent.

### 4.1 Agents and Devices ID

Identification of both ends (agents and devices) are implemented by using a hash function over MAC address and IP address. IPv4 and IPv6 address are supported. When a device contact to agents uses the IP address and MAC hash has ID. Device use this number (currently 160 bits) as ID before logs can be transported form devices to agents. This ID doesn't change. With this approach a device can be always identified even if change its IP address or MAC.

**Figure 3 An Agent waiting for connections.**

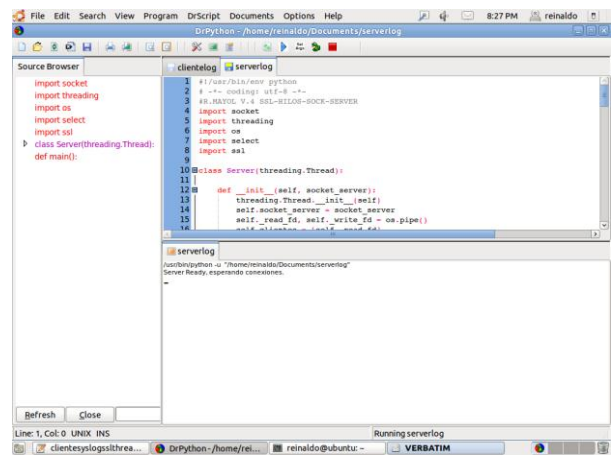

### 4.2 Agents and Devices Authentication.

Both agent and device have a Digital Certificate (X.509 v3) and a private key. Therefore, both ends can be authenticated using a challenge protocol. Before an agent, or a device, can exchange data, both must obtain the digital

---
[7] Syslog only have 23 messages models. Some of them, like UUCP are obsoletes.

certificate of the other from a Certificate Authority. This process is not part of the protocol.

When a device wants to send data to an agent generates a challenge (consisting of a 4096-bit random number). This challenge is encrypted by using the device´s private key and send to agent. The agent decrypts the challenge (using the device's public key) and encrypts it again but with its own private key. The device now decrypts the challenge using the agent´s public key. If the received value is identical to the generated both ends are correct. (see Figure 5)

This process is now implemented using RSA protocol. The key and certificates can made using SSL.

**Figure 4 Two devices establishing an SSL tunnel with the agent after performing the authentication.**

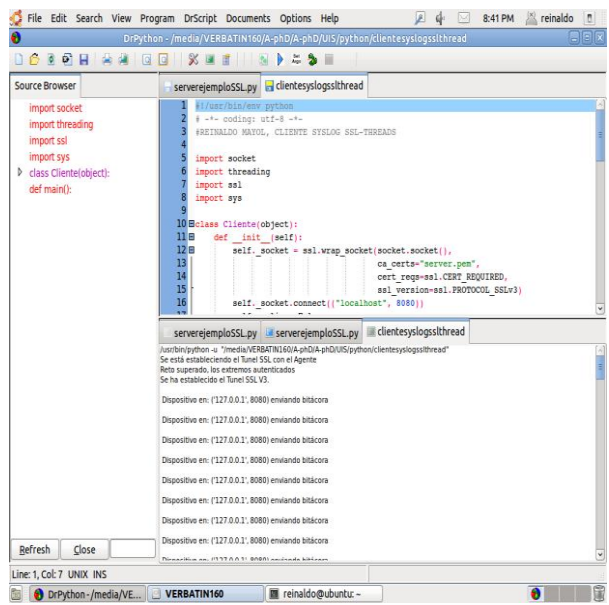

### 4.3 Data Confidentiality

The confidentiality of communications between agents and devices is an important element. We have implemented the protocol Rinjdael with keys of 128 bits. A session key is exchanged using Diffie-Hellman.

To prevent that a data packet could be captured and repeated indiscriminately, session keys has time to live (TTL). The TTL is a value in Kilobits. When a key was used to encrypt more than the amount of data defined by the TTL, it should be changed. At that point the device and the agent must negotiate a new session key using Deffie-Hellman protocol. Now we are working in a window´s system. If a packet was received then other packets with similar number are rejected.

### 4.4 Data Integrity

The data integrity must be preserved. To accomplish this, the messages will be digitally signed. The version of the agents currently implemented using SHA -2.

We are currently testing an alternative that would consume fewer resources, especially bandwidth. This is to authenticate communications groups instead of individual packages. The transmitter (devices or agents) generates the signature after that an amount of data transmitted. If the signature is incorrect all the affected packages should be retransmitted.

### 4.5 Agent Discovery and Selection

The devices must define the agent that receives information (logs). Devices can discovery agents in the same network.

To discover agents, devices use the following protocol:

1. Devices send multicast messages and send them an Agent-Discovery Message (ADM)[8].
2. When an agent receives the ADM message, it sends back a Communications Offer message (AOM)[9] to device.

A device can receive offers from many agents therefore; the devices could select the agent to send information. To do it, devices use following protocol:

1. Select the agent used before.
2. If the device has never send data before, select an agent on the same network.

---

[8] ADM is a single message with device ID. This message is digitally signed.
[9] AOM is a single message with agent ID. This message is digitally signed.

3. If two or more agents are available on the same network one of them is randomly selected.
4. If there are no agents in the same network or no one sent offers, another agent can be used, but must be manually configured.

## 5. Tests on agents

The tests were performed with 8 machines running linux (debian) as devices and an agent (also debian).

Each device transmitted 200 Mb from its real log. Figure 5 show various devices communicating witch an agent during test.

**Figure 5. Various devices transmitting data to an agent using Ipv6 address.**

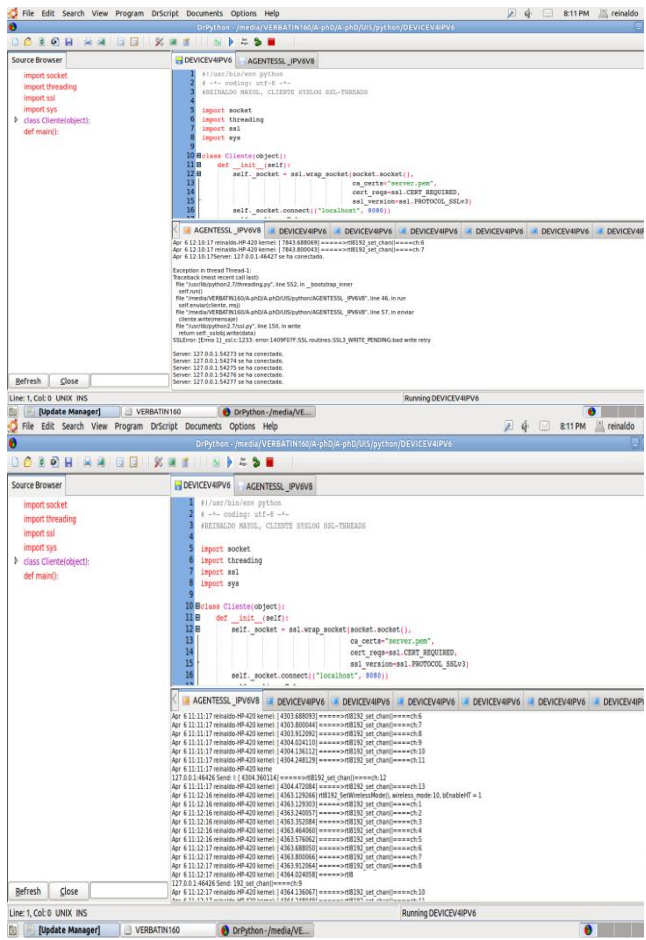

As result of the tests we found that performance drops significantly when number of devices simultaneously connected increases with the same agent. This can be caused by the selection of programming language. This is an expected result because the degree of development of the project and the objectives in this stage.

Another test made was to capture and repeat a packet indiscriminately in the network. It was confirmed that damage was limited to the period of validity of the session keys.

The tests were conducted using both IPv4 and IPv6 (in local-link mode). There were no significant differences in performance between the two tests.

Finally, tests were performed using decoys and false devices and agents. In all cases the tests showed that the connections from the apocryphal elements were rejected.

## 6. Conclusions and Future work

The information described in this article is part of a larger project designed to create a secure distributed environment of e- analysis. Here we have described only the proposed design of agents capable to obtain information from devices and pre-processing it.

The goals proposed for this stage of development were achieved. We have obtained a prototype of a functional agent capable of receiving data from remote devices safely and reliably. This prototype can solve some of the current deficiencies in management systems logs.

At this stage the tests were satisfactory and showed the feasibility of the proposed prototype.

Future work in agents must include:

- Ipv6 anycast communications between agents and devices as part of a fault tolerance mechanism.
- Agents auto-discovery process.
- Orchestration process implementation and communications from agents to devices.
- Data compressing
- Data normalization using XLM format
- Addition of metadata to describe data quality
- Implement a windows system to reduce the impact of packet repetition.

Additionally, the orchestration process, the transportation protocol between agents and servers and system testing scheme should be implemented and tested in the future.